\documentclass[prb,twocolumn,showpacs,preprintnumbers,amsmath,amssymb]{revtex4}
\usepackage{graphicx}
\usepackage{dcolumn}

\begin{document}

\title{Effects of shadowing in oblique-incidence metal (100)  epitaxial growth}

\author{Yunsic Shim}
\email{yshim@physics.utoledo.edu}
\author{Jacques G. Amar}
\email{jamar@physics.utoledo.edu}
\affiliation{Department of Physics \& Astronomy \\ University of Toledo,  Toledo, OH 43606}

\date{\today}

\begin{abstract}  
The effects of shadowing in oblique incidence metal (100) epitaxial growth are studied using a simplified model. We find that many of the features observed in 
Cu(100) growth,  including the existence of a transition 
from anisotropic mounds to ripples perpendicular to the beam, can be explained purely  by geometrical effects.   We also show that  the formation of (111) facets 
 is crucial to the development of ripples  at large angles of incidence. 
 A second transition to   `rods'  with (111) facets oriented parallel to the beam is also found
 at high deposition angles and film thicknesses. 

\end{abstract}
\maketitle

While a variety of  surface  relaxation processes, such as adatom diffusion on terraces and near steps, as well as edge-diffusion and detachment, are usually assumed to  determine  the 
surface morphology  in epitaxial growth, recently it has been shown\cite{Dijken,Dijken2,voter,Yuamar1}  that the deposition process    can also play an important role. 
For example,  in the case of unstable metal epitaxial  growth with  an Ehrlich-Schwoebel (ES) barrier\cite{ES}  to diffusion over descending steps,   
the short-range (SR) attraction of depositing atoms to the substrate can significantly enhance the surface roughness and selected mound angle.\cite{Yuamar1}  
In addition, for  sufficiently large angles of incidence, the deposition angle can also   play an important role. For example, in recent experiments  on  Cu/Cu(100) epitaxial growth\cite{Dijken,Dijken2}  
at $250$ K a gradual transition was observed  from symmetric mound structures  for deposition angles up to $\theta = 55^o$ (where $\theta$ is the angle between the beam and the substrate normal), to asymmetric mounds with increasing   slopes for deposition angles up to $70^o$,  to asymmetric  ripples   oriented perpendicular to the beam with (113)/(111) facets on the shadow/illuminated sides  at $\theta = 80^o$.   
Similar results have been obtained in grazing incidence Co/Cu(001) growth \cite{Dijken3} for which it was found that the resulting surface  anisotropy also  leads to  strong
 uniaxial magnetic anisotropy. Thus, understanding the effects of oblique incidence deposition 
is important since it may  lead to the possibility of controlling both 
the surface morphology and magnetic  properties in epitaxial growth.

Although the effects of shadowing on   thin-film morphology   
have  been extensively studied   in the case of amorphous and polycrystalline columnar growth,\cite{gladrefs} the case of epitaxial growth is not as well understood. 
We note that in Ref.~\onlinecite {Dijken2}  it was shown  that 
at large angles of incidence 
the  long-range (LR)   van der Waals  attraction of depositing atoms to the substrate
may play an  important role in addition to the SR attraction. 
Therefore, a fully realistic simulation  of oblique-incidence epitaxial growth
can be very time-consuming since it must take into account both the LR and SR interactions. 
As a result,  recent simulations  have focused  on the submonolayer regime\cite{Seo} or at most on the first few layers of vicinal growth.\cite{Seo2}  
However, at high angles of incidence the effects of shadowing and crystal geometry also play an increasingly important role.  Therefore,  it is of interest to determine to what extent these purely geometric effects 
may determine the surface morphology.

Here  we present the results of simulations carried out using a simplified model of 
fcc(100)  epitaxial growth in which the  effects of  shadowing 
are included  but  not  the additional modifying effects of the SR and LR attraction.  
Our results indicate that many of the qualitative and semi-quantitative features observed in 
Cu(100) growth, \cite{Dijken, Dijken2} including the existence of anisotropy in the submonolayer regime, can be explained purely by geometrical (shadowing) effects. 
In addition, we find that the formation of (111) facets is crucial to the formation of ripple structures at large angles of incidence.  The dependence of the critical thickness for ripple formation on deposition angle is also studied and good scaling behavior  is obtained for the correlation length perpendicular to the beam  as a function of deposition angle and film thickness.  
At higher thicknesses we also observe a second transition to `rods' with (111) facets oriented parallel to the beam.   
We also find that the surface width   increases exponentially with deposition angle and exhibits excellent scaling as a function of film thickness.

Except for   deposition, our model is   very similar to previous   models\cite{Amarprb,Shimprb} used to study   metal (100)  growth at normal incidence 
in which the correct crystal geometry has been taken into account.  
In particular, atoms are   deposited with a (per site) deposition rate $F$, while adatoms (monomers) on a flat terrace are  assumed to diffuse with hopping rate $D$.  Since the ES barrier typically plays an important role in metal epitaxial growth, the rate for an adatom at a descending step-edge to diffuse over the step is   given by $D_{ES} = D e^{-E_{ES}/k_B T}$ where $E_{ES}$ is the Ehrlich-Schwoebel barrier.  
Compact islands are also assumed and accordingly a moderate amount of edge- and corner-diffusion 
is also included,    while the  attachment  of atoms to   existing islands is   assumed to be irreversible. Thus the most important parameters in our model are the deposition angle $\theta$,  the ratio $D/F$ of the monomer diffusion rate to the deposition rate, and the magnitude of the ES barrier.

To take into account the effects of shadowing and crystal geometry, 
each deposited atom is assumed to travel ballistically   until  its   distance to the closest substrate atom  is less than or equal to the nearest-neighbor distance.  
The depositing atom  then `cascades'  randomly via downward funneling (DF) \cite{df}  from a site corresponding to this atom  until it reaches a four-fold hollow site.
Thus, in our model  atoms deposited on (111) facets are assumed to diffuse essentially instantaneously  to the terrace below. Given the extremely low barriers for diffusion on metal (111) surfaces (approximately $0.05$ eV for Cu(111)\cite{111barrier}) for temperatures which are not too low this is a very reasonable approximation.

In most of our simulations two different deposition rates were used - one corresponding to a `slow' deposition rate  ($D/F = 10^5$) and the other corresponding to a `fast' deposition rate ($D/F = 5000$).  Similarly, two different values of the ES barrier were also used, one corresponding to a `moderate' barrier at room temperature ($E_{ES} = 0.07$ eV) and the other corresponding to a large, effectively infinite ES barrier.  As in the experiments of van Dijken et al \cite{Dijken, Dijken2} 
the overall deposition rate was assumed to be independent of deposition angle.  
 Similarly,  the azimuthal angle  was chosen such that the deposition direction was  parallel to the close-packed step-edge, i.e. along the [110] direction.

\begin{figure}[  t] 
\includegraphics[width=6.4cm] {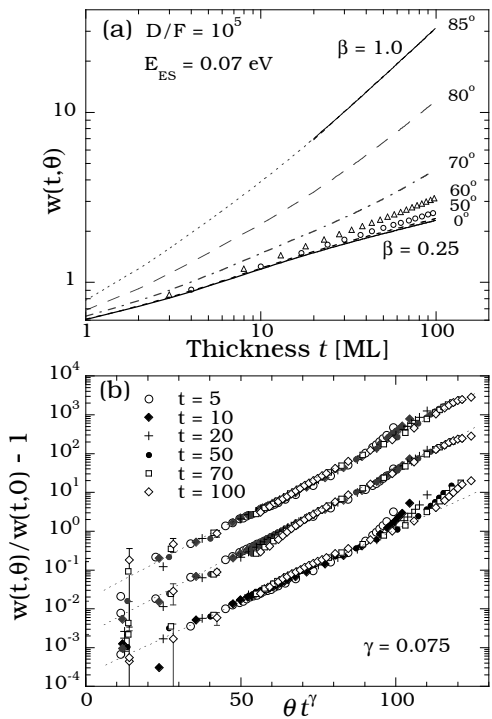}
\caption{\label{Fig1}  {(a) Surface roughness  (in ML)
as function of film thickness 
for $\theta = 0^o - 85^o$ for case of slow deposition 
and moderate ES barrier 
(b) Relative width deviation $\delta w/w(t,0)$ 
as function of scaled deposition angle $\theta t^\gamma$ for $\theta = 10^o - 88^o$ and $t  = 5 -100$ ML for three different cases:  fast deposition, high ES barrier (top),  fast deposition moderate ES barrier (middle),  and slow deposition, moderate ES barrier (bottom).   }} 
\end{figure}

In order to understand the dependence of the surface morphology 
on deposition conditions we have calculated a variety of different quantities 
as a function of average film thickness $t$ (where $t$ is in ML) and deposition angle $\theta$.  
These include the r.m.s. surface height or ``width" $w$, the  lateral correlation lengths   $\xi_{\|}$ and $\xi_{\perp}$ determined from the zero-crossing of the  height-height correlation functions   parallel and perpendicular  to the beam, 
and the anisotropy $\alpha = \xi_{\perp}/\xi_{\|}$. 
Our simulations were typically carried out using relatively large system sizes ($L = 512$) and averaged over 100 runs.

Fig.~1(a)  shows typical results  for the surface width 
as a function of film thickness  for deposition angles ranging from $0^o$ (normal incidence)  to $85^o$, for the case of slow deposition and a moderate ES barrier.  
As in the Cu(100) growth experiments of van Dijken et al,\cite{Dijken} for  $\theta \le 50^o$  the effects of oblique incidence and   shadowing  on the surface roughness are relatively weak.  However, for larger  deposition angles 
the surface roughness increases significantly with increasing deposition angle.  As a result, the value of the effective roughening exponent  $\beta$  at large film thickness $t$  (where $w \sim t^\beta$) increases from $\beta \simeq 1/4$ 
for small angles 
to a value close to $1$ at $\theta = 85^o$.   

Fig.~1(b) shows the  corresponding results for the relative deviation  in the width $w(t,\theta)$ 
compared to the width at normal incidence 
as  a function of the scaling variable $\theta t^\gamma$, 
where the scaling exponent $\gamma \simeq 0.075$.    
Also included are similar results obtained for the cases of fast deposition with a moderate ES barrier, and  fast deposition with a large ES barrier, 
which have been shifted up by factors of $10$ and $100$ respectively for clarity. 
As can be seen, there is excellent scaling 
for all three sets of deposition conditions.  In addition, we note that the (unshifted) scaling functions in all three cases are almost   identical except for some small deviations for large $\theta t^\gamma$.
The linearity of the scaling function in each case  over approximately $4$ decades    shows 
that for fixed deposition rate 
the relative width deviation increases exponentially with increasing deposition angle $\theta$. 
The small value of the scaling exponent $\gamma$ also  indicates 
that the  surface width depends much more strongly on deposition angle than on film thickness.

\begin{figure}[t] 
\includegraphics[width=8.0cm] {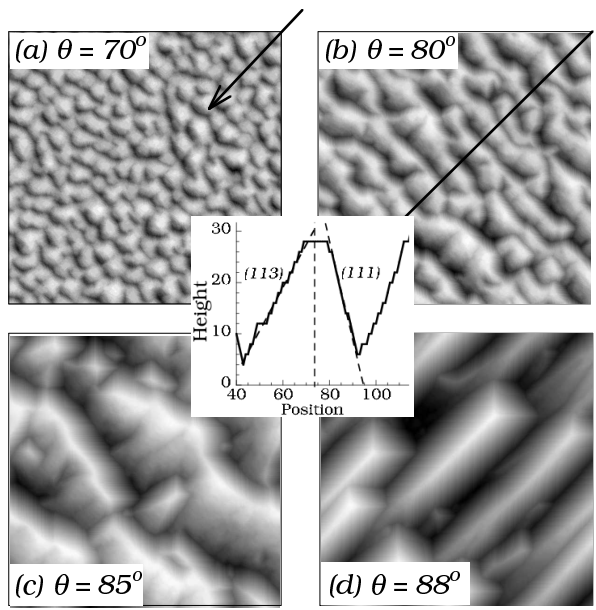}
\caption{\label{Fig2}  {Gray-scale pictures of surface morphology (system size $L = 512$) for same conditions as in Fig. 1(a) at $t = 50$ ML.  Arrow indicates deposition direction.  Inset shows ripple profile taken along line in (b). }} 
\end{figure}

\begin{figure}[] 
\includegraphics[width=8.0 cm] {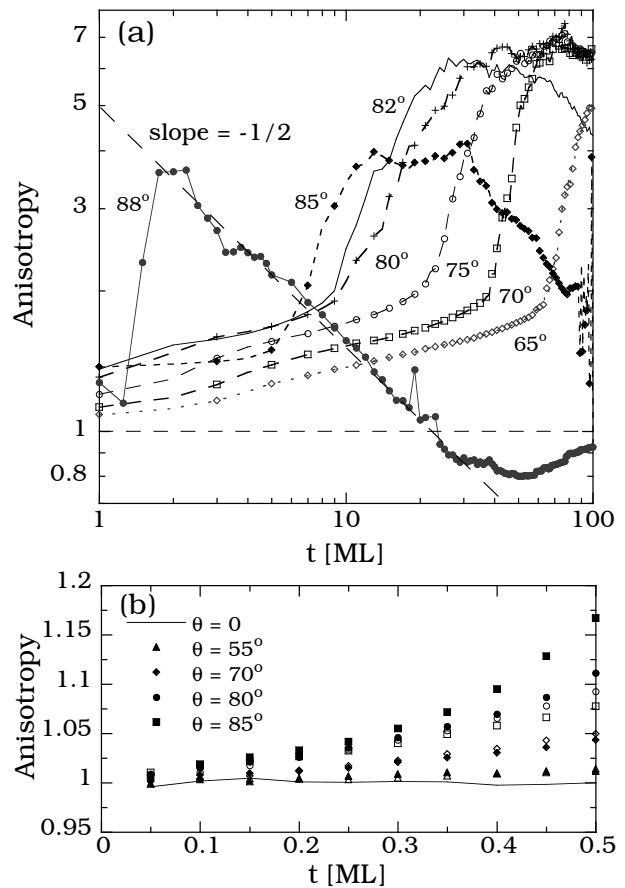} 
\caption{\label{Figrc2orc1}  {Anisotropy $\alpha = \xi_\perp/\xi_{\|}$ as a function of film thickness   for the case of slow deposition with a moderate ES barrier.  (a) multilayer regime  (b) submonolayer regime. In (b) the corresponding results for a large ES barrier are also shown  (open symbols). }}
\end{figure}

 Fig.~2 shows   typical pictures of the  surface morphology   at  a film thickness  
 of  $50$ ML
 for different  deposition angles for the case of slow deposition (moderate ES barrier). 
 At a deposition angle $\theta \simeq 70^o$   (Fig.~2(a))   shadowing leads to a   preference for mounds to coalesce  along the direction perpendicular to the beam and the onset of ripple formation. 
  At a somewhat larger angle ($\theta \simeq 80^o$)  
 asymmetric ripples 
  with (111) facets on the illuminated side and (113) facets on the shadow side are formed (see Fig.~2(b) and inset)  as in the experiments of Dijken et al.\cite{Dijken, Dijken2}  The asymmetry of these ripples can be more clearly seen at $\theta \simeq 85^o$ (Fig.~2(c)) which shows clearly the (111) facets on the illuminated side.     We note that at this angle 
  there is also evidence for 
 a competition between the growth of asymmetric ripples perpendicular to the beam direction, and the growth of  `rods' with (111) facets parallel to the beam. 
 In simulations  with reduced corner diffusion this competition leads to an extended regime in which  isotropic pyramidal structures with (111) facets on all four sides are formed, similar to what is observed   experimentally in Ref.~\onlinecite{Dijken}.  
  Finally, at $\theta = 88^o$ (Fig.~2(d)) the perpendicular ripples are completely replaced by the growth of  `rods'  with (111) facets growing parallel to the beam. 
 
 The dependence of the anisotropy on both deposition angle and film thickness for this set of deposition conditions can be seen more quantitatively  in Fig.~3.   
  At a critical thickness $t_c$, which increases rapidly  with decreasing deposition angle,  there is an abrupt  {\it increase}  in the anisotropy,  
 corresponding to the onset of ripple formation.  The anisotropy  then saturates as the ripples continue to elongate as well as coarsen  but  then decreases as   $t^{-1/2}$ at  higher film thicknesses. 
 As can be seen in Fig.~3(b), 
 at large angles there is already a small but noticeable anisotropy in the submonolayer
 regime, 
 even in the absence of steering effects  
 due to SR and LR attraction.\cite{footnote} 
  This anisotropy 
 is primarily due to the fact that shadowing tends to  inhibit (enhance) the coalescence of islands along the directions parallel (perpendicular)  to the beam.

 A detailed examination of the surface morphology 
 in the multilayer regime 
  indicates that the abrupt increase of the anisotropy and formation of ripples is due to   the formation  of  (111) facets on the illuminated sides of mounds.  Since  (111) facets can efficiently capture and transport depositing atoms to the sides,  this leads to a   strong enhancement of mound coalescence in the direction perpendicular to the beam, 
 followed by the formation and growth of 
  ripples with extended  (111) facets on the illuminated side.    However, the saturation and eventual decrease of the anisotropy   may be attributed to the formation of 
   structures with (111) facets on the sides {\it parallel to the beam} as well as on the illuminated side (see Fig.~2(d)) whose growth competes with ripple growth.  Such structures are particularly stable at large deposition angle, while  due to the large flux on the illuminated facet 
 they  tend to grow linearly in the direction of the beam.  
 The decay of the anisotropy with exponent $-1/2$ (see Fig.~3(a)) 
 may then be explained by the fact that in this regime the correlation length perpendicular to the beam grows as $\xi_{\perp} \sim t^{1/2}$ 
 while the correlation length parallel to the beam 
 grows linearly with film thickness, e.g. $\xi_{\|} \sim t$.

\begin{figure}[t] 
\includegraphics[width=8.0cm]  {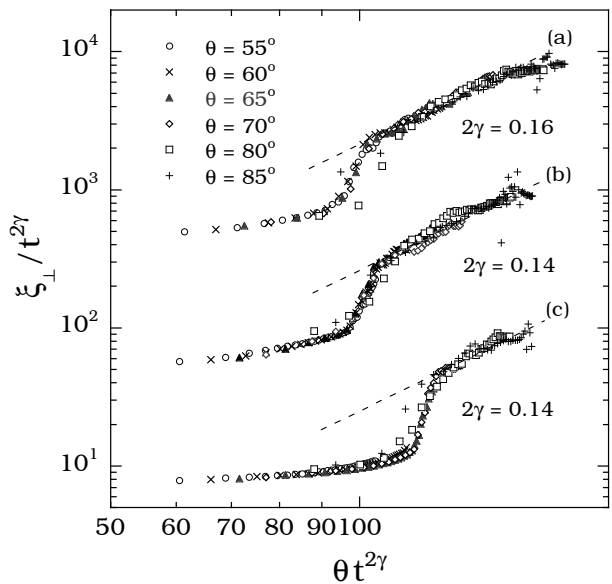}  
\caption{\label{Figrc2}  {Scaled 
correlation length $\xi_\perp/t^{2\gamma}$  as function of scaled deposition angle for three different sets of deposition parameters:  (a) fast deposition, large ES barrier (b) fast deposition, moderate ES barrier, and (c) slow deposition, moderate ES barrier.  Slope of dashed lines is approximately $2.6$. }} 
\end{figure}

    Fig.~4  shows   the scaled perpendicular correlation length  as a function of   scaled deposition angle $\theta t^{2\gamma}$ where  $\gamma \simeq 0.075$, for $t = 5 - 100$ ML and $\theta = 55^o - 85^o$ for three different sets of deposition parameters (see caption) where (a) and (b) have been shifted upwards by factors of $100$ and $10$ for clarity. 
As can be seen, 
  in each case there is excellent scaling     over a wide range of film thicknesses and deposition angles.  
  This  scaling behavior also indicates that ripple formation may be expected to occur  for deposition angles as low as $55^o$,  although the  relatively low value of $\gamma$   indicates  that the critical thickness   for ripple formation increases rapidly with decreasing deposition angle.  
We note that the slope of the scaling function after the  onset of ripple formation (dashed lines in Fig.~4) 
   corresponds to  the $t^{1/2}$ growth of   ripples with (111) facets perpendicular to the beam,  
   and is consistent with domain growth with non-conserved order parameter, \cite{Bray} 
  while the  later  `saturation'   
corresponds  to the slower coarsening of  structures with (111) facets parallel to the beam, 
and  is consistent with one-dimensional fluctuation induced coarsening.\cite{Tang}
We also find excellent scaling for the scaled correlation length $\xi_\perp/l_d$ as a function of the scaled film thickness $t/l_d^2$ where $l_d \simeq (D/F)^{1/6}$   is the diffusion length, although the resulting scaling function depends  on deposition angle.

Finally, we discuss the factors that may affect ripple formation in oblique-incidence epitaxial growth. 
Since ripple formation is associated with the formation of (111) facets on the illuminated sides of mounds, 
we expect that anything that promotes this 
tends to enhance ripple formation. 
For example, since a large ES barrier leads to faster mound formation as well as a larger selected mound slope, increasing the ES barrier leads to earlier ripple formation as shown in Fig.~4.  Similarly, decreasing the flux and thus increasing $D/F$,  
or increasing the rates of edge- and/or corner-diffusion 
delays the formation of (111) facets and ripples  as again shown in Fig.~4.  
We have also considered the 
case of  low-temperature growth ($D/F \simeq 0$) but assuming  that DF as well as fast diffusion on (111) facets remain active.   Somewhat surprisingly, we find that even in this case, for which
the surface is `stable' at normal incidence,\cite{Amarprb}  
  the formation of well-defined ripples with (111) facets 
is observed 
at large deposition angles ($\theta \ge 60^o$). 
Thus,  we conclude that while ripple formation may be enhanced by the presence of a mound instability, 
 it is primarily a geometric effect due to   shadowing as well as the existence of rapid diffusion on (111) facets.   

In conclusion,  we have used a simplified model  to study  the evolution of the surface morphology and anisotropy in oblique-incidence metal (100) growth.  Our results indicate that much of the qualitative and even semi-quantitative behavior can be explained  by geometric effects which dominate at large 
deposition angles. 
Our results also indicate the existence of a competition between ripples perpendicular to the beam and structures with (111) facets parallel to the beam which eventually leads to a novel transition from perpendicular ripples to `rods' at  high deposition angles and film thicknesses.  
While we expect that the inclusion of SR and LR attraction will   affect    the results presented here, the qualitative picture is unlikely to change.  
In particular, due to the extreme sensitivity on deposition angle, the most significant effects of including such interactions are  likely to be a slight shift in the angles at which ripple and/or rod formation take place as well as  a slight modification of the anisotropy.  It   is also possible that the inclusion of the SR and LR attraction may   further stabilize the ``isotropic phase" corresponding to the competition between rods and ripples as is observed experimentally. 
In  future work, we plan  to carry out   multiscale simulations in order to determine in more detail to what extent the effects of SR and LR attraction may modify this picture.

\begin{acknowledgments}

This research was supported by grants from the Petroleum Research Fund and   NSF as well as by a  grant of computer time from the Ohio Supercomputer Center.
\end{acknowledgments}


\begin{thebibliography}  {99}
\bibitem{footnote} 
In the submonolayer regime the anisotropy was estimated by calculating the ratio of the average island lengths perpendicular and parallel  to the beam respectively.
\end{thebibliography}
 \end{document}